\documentstyle[aps,floats,prb]{revtex}
\input psfig
\begin{document}
\twocolumn[\hsize\textwidth\columnwidth\hsize\csname @twocolumnfalse\endcsname

\title{
Semiempirical Hartree--Fock calculations for KNbO$_3$
}
\author{R.~I.~Eglitis,\cite{*} A.~V.~Postnikov, and G.~Borstel}
\address{
Universit\"at Osnabr\"uck -- Fachbereich Physik,
D-49069 Osnabr\"uck, Germany}
\date{Received 31 January 1996; revised manuscript received 10 April 1996}
\maketitle
\begin{abstract}

In applying the semiempirical intermediate neglect of
differential overlap (INDO) method
based on the Hartree-Fock formalism to a
cubic perovskite-based ferroelectric material KNbO$_3$,
it was demonstrated that the accuracy of the method
is sufficient for adequately describing the small
energy differences related to the ferroelectric instability.
The choice of INDO parameters has been done for a system
containing Nb. Based on the parametrization proposed,
the electronic structure, equilibrium ground state structure
of the orthorhombic and rhombohedral phases,
and $\Gamma$-TO phonon frequencies
in cubic and rhombohedral phases of KNbO$_3$ were calculated
and found to be in good agreement with the experimental data
and with the first-principles calculations available.
\end{abstract}
]
\section{Introduction}
\label{sec:intro}
Potassium niobate, a perovskite-type ferroelectric material
isostructural to barium titanate,
has been subject to numerous {\em ab initio}
electronic structure calculations during recent years.
Earlier calculations have been performed for the ideal cubic
perovskite structure in order to obtain electron band
structure and to interpret optical\cite{olcao}
or x-ray photoelectron\cite{xps} spectra.
Since then, special attention has been paid
to the total energy calculations
making it possible to determine
the equilibrium geometry\cite{ktn3,ksv94},
phonon frequencies\cite{sb92,phonon,loto,chain},
and interatomic interaction parameters
defining the ferroelectric phase transitions\cite{ksv94,phase}.

Most of the calculations cited have been performed
using the local density approximation (LDA), either
with the pseudopotential method (Refs. \onlinecite{ksv94}
and \onlinecite{loto}) or with linearized augmented plane wave
(Refs. \onlinecite{sb92}, \onlinecite{chain} and \onlinecite{s95})
or linearized muffin-tin orbital (LMTO, Refs. \onlinecite{ktn3}
and \onlinecite{phonon}) method. The latter two approaches
use some (different) forms of series expansions for the potential inside
the muffin-tin spheres and in the interstitial.
In the LDA-based calculation schemes, the potential is local
and orbital independent, unless Coulomb correlation effects
are {\em ad hoc} introduced by one or another implementation
of self-interaction corrections,\cite{sic} or within
the LDA+$U$ (Ref. \onlinecite{ldau}) formalism. This seems to be
especially important for treating localized states,
such as, e.g., those of transition-metal impurities in insulators.
As has been shown in Ref. \onlinecite{mgo} for Fe in MgO,
the straightforward implementation of a LDA scheme
may lead to wrong results with respect to the energy positioning
of impurity levels and the magnetic moment related to the impurity.

In contrast to LDA, the Hartree-Fock formalism
automatically incorporates
the dependence of the potential on the symmetry of a particular
orbital, as well as on whether this orbital is occupied or not.
Another convenient property of the Hartree-Fock formalism
that is typically realized on a tight-binding basis set
is that it can be more or less directly applied to
crystal surfaces, providing correct asymptotics
of the electron density at the vacuum side.

{\em Ab initio} Hartree-Fock calculations are
excessively computationally demanding (in the sense that
quantitative results of comparable, or better, accuracy
are in many cases obtainable within the LDA at much
lower cost), therefore the applications to perovskite
systems are not numerous. As an eventually single example,
finite cluster Hartree--Fock calculation has been
reported for a fragment of KNbO$_3$ structure.\cite{donner}
However, a simplification on top of the Hartree--Fock method
known as intermediate neglect
of the differential overlap\cite{indo1,indo2} (INDO)
lets to decrease the computational effort considerably,
at the price of treating several parameters as
fitting variables, to be defined from outside the
calculation scheme. In contrast to model calculations,
which usually require an {\em ad hoc} fitting,
the INDO parameters are believed to be largely
transferable, so that, once determined for some
chemical constituent, they may be successfully applied
in the calculations for a variety of chemical substances
where the latter participates.

Typical fields of INDO applications include
various defects systems based on silica,\cite{indo1,indo3}
ionic oxides, such as MgO,\cite{indo1,indo2}
corundum,\cite{coru1,coru2} zirconia,\cite{zro2}
or alkali halides.\cite{indo2,lif} The choice of INDO parameters
is not a straightforward procedure but rather a trial-and-error
loop, aimed at reproducing reasonably well band structure,
equilibrium geometry, and characteristic energy
differences for molecules or crystals, as based on
experimental measurements or {\em ab initio} calculations.
The list of parameters for several elements is given
in Refs. \onlinecite{indo1} and \onlinecite{indo2}
along with some discussion on the parameter optimization
for ionic crystals.

The aim of the present paper is to demonstrate
that the semiempirical INDO method may work well for
perovskite-type ferroelectrics, and to provide
the optimized set of INDO parameters
for all constituents of KNbO$_3$.
Perovskites are generally expected
to present a problem for any parametrized method,
because of varying degree of covalency depending
on chemical composition and because of strong
polarizability of transition metal--oxygen bonds.
An additional difficulty related to ferroelectric perovskites
is that the energy differences that play a role
in stabilizing the ferroelectric distorted structure,
due to a fine balance between long-range Coulomb forces
and short-range chemical bonding, have the order of magnitude
of 1~mRy per formula unit or smaller,
i.e., much lower than $\sim$1~eV energy differences being
discussed, e.g., in relation to charged defects in silica.\cite{indo3}
Since this is the first, to our knowledge, application
of the INDO method to perovskite systems,
the question is to be answered whether the accuracy
of the parametrized INDO method is sufficient
to describe the ferroelectric instability, and whether
the description of the underlying energetics is reliable.
We optimize the INDO parameter set based on the
comparison with available {\em ab initio} calculation results
and experiments, and answer the above question positively
by presenting the INDO calculations for atomic displacement
patterns and phonon frequencies
that are in good agreement with experimental data.

The paper is organized as follows. In Sec. \ref{sec:method},
we describe the essential features of the INDO method
and the meaning of underlying parameters.
In Sec. \ref{sec:param}, the choice of the INDO parameters
used in our calculation is specified,
based on the comparison with total energy
{\em ab initio} calculations and the experimental
structure data for KNbO$_3$.
In Sec. \ref{sec:results},
atomic coordinates in the room-temperature orthorhombic phase
and in the low-temperature rhombohedral phase
are found by the total-energy-based structure optimization,
and the results of INDO calculations for $\Gamma$-TO
phonon frequencies are discussed.

\section{INDO method and parameter optimization}
\label{sec:method}
The calculation scheme of the Hartree-Fock-Roothaan method
in the INDO approximation is discussed in detail
in Refs.\onlinecite{indo1} and \onlinecite{indo2}.
Basically, the procedure
reduces to diagonalizing the matrix of the Fock operator
to get the one-electron energies, and the linear combination
of matrix elements with appropriate weights, depending
on the occupation of corresponding one-electron states,
provides the total energy.
The fixed basis set is minimal in the sense that
each of the atom-centered functions
related to the valence-band states (4 in total per oxygen atom,
9 per transition-metal atom) is encountered only once.
The construction of the on-site and off-diagonal
parts of the Fock matrix is determined in terms of
several empirical parameters, labeled by the atom type $A$ and
the index of the atomic orbital (AO) $\mu$ (see Ref.\cite{indo1}).
The interaction of an electron
in the $\mu$th valence AO on atom $A$ with its own core
\[
U_{\mu\mu}^A=-E^A_{\text{neg}}(\mu)-\sum_{\nu\in A}
  (P_{\nu\nu}^{(0)A}\gamma_{\mu\nu}-
  \frac{1}{2} P_{\nu\nu}^{(0)A} K_{\mu\nu})
\]
contains, apart from the $\zeta_{\mu}$ value, which specifies
the Slater exponent for a one-exponential basis function
and hence Coulomb and exchange integrals $\gamma_{\mu\nu}$
and $K_{\mu\nu}$,
the initial guesses for the diagonal elements
of the density matrix $P_{\mu\mu}^{(0)A}$
and for the energy of the $\mu$th AO $E_{\text{neg}}^A(\mu)$,
i.e., the ion's electronegativity.
The interaction of the $\mu$th AO with the core
of another atom $B$ is approximated as
\[
V_{\mu}^B=Z_B \left\{ 1/R_{AB} +
	\left[ \langle\mu\mu|\nu\nu\rangle - 1/R_{AB} \right]
        \exp(-\alpha_{AB}R_{AB}) \right\},
\]
where $R_{AB}$ is the distance between atoms $A$ and $B$,
$Z_{B}$ is the core charge of atom $B$,
and parameter $\alpha_{AB}$ describes the non-point character of this
interaction.

Finally, the resonance-integral parameter $\beta_{\mu\nu}$ enters the
off-diagonal Fock matrix elements for the spin component $u$:
\[
F_{\mu\nu}^u=\beta_{\mu\nu}S_{\mu\nu}-P_{\mu\nu}^u
  \langle\mu\mu|\nu\nu\rangle,
\]
where the $\mu$th and $\nu$th AO are centered at different atoms,
$S_{\mu\nu}$ is the overlap matrix between them,
and $\langle\;|\;\rangle$ are two-electron integrals.
Parameters $\zeta_{\mu}$, $\beta_{\mu\nu}$, $\alpha_{AB}$
and $E^A_{\text{neg}}(\mu)$ are usually fixed throughout
the iterations, whereas $P_{\nu\nu}^{(0)A}$ may be corrected
as the self-consistency is being achieved.

It is in principle possible to implement the calculation
in such way that the diagonalization is done for a number
of {\bf k} vectors per iteration. However, conventional
usage of the INDO method, given the low symmetry and possibly the
lack of translation invariance in the systems it is usually
applied to, restricts the diagonalization to the $\Gamma$ point
of the Brillouin  zone only, taking instead a supercell,
or large unit cell (LUC), all atoms of which contribute
to the expanded basis set. For ideal systems, the enlargement
of the unit cell is equivalent to increasing the density
of the {\bf k} mesh in band-structure calculations, since
the $\Gamma$ point of the reduced (in the supercell) Brillouin zone
maps onto different points of the original Brillouin zone
of the single cell. For defect systems, there is no problem
to treat, e.g., discrete impurity states in the dielectric gap,
if one or few impurity atoms are included along with tens
of bulk atoms in the LUC, since the {\bf k} dispersion
of such states is negligible. Anyway, the enlargement
of the unit cell in case of defect systems
increases the variational freedom of the basis set.

Since the construction of the parametrized Fock matrix
is straightforward, the bottleneck of the method
in what regards the performance and accuracy
is the diagonalization of large matrices.
Compared to precise LDA-based schemes, such as, e.g.,
full-potential (FP) LMTO, which usually employ
multiple-tail representation of basis functions,\cite{msm1,sav}
the INDO method manages to handle considerably
larger supercells. Compared to efficient
minimal-basis computational schemes as, e.g.,
tight-binding LMTO,\cite{tblmto} INDO may exhibit
such advantages as unrestricted spatial form of the potential,
absence of muffin-tin boundary conditions, and of space-packing
empty spheres.

As a method essentially based on the Hartree-Fock approximation,
INDO does not provide a convenient option to incorporate correlation
effects into one-electron equations, as may be to some extent done
within the LDA by an appropriate choice of the exchange-correlation
potential. As a result, the dielectric band gap comes out
in INDO overestimated usually by 3--5 eV (see, e.g.,
Ref.\onlinecite{indo1} and \onlinecite{coru1}).
Moreover, the lack of correlation effects, which imply
some additional repulsion between electrons, overestimates
the chemical binding and results in even more underestimated
equilibrium bond lengths than is known to be the case
in LDA calculations. Some part of correlation corrections
(usually referred to as short-range corrections
as they are mostly of intra-atomic nature)
may be, however, incorporated in the choice of INDO parameters,
since the latter are based on experimental or other external
information anyway, and this may to some extent
improve the two shortcomings mentioned.

\section{Parameter optimization}
\label{sec:param}
In the choice of INDO parameters for our calculation, we
relied on the experimental information available and
on the data of {\em ab initio} calculations for KNbO$_3$
(cited in Sec. \ref{sec:intro}), which essentially
agree in the description of the band structure.
Whereas INDO calculations for many oxides and potassium
salts have been done earlier, and
the corresponding parameters for O and K tabulatedi,\cite{indo2}
no INDO calculations involving Nb have been, to our knowledge,
done by now, so one-center and all involved two-center
parameters had to be found.
Since $E_{\text{neg}}$ is related to the central energy position
of an AO in question which is hybridized with many
other states throughout the valence band, we calculated
partial density of states (DOS) by sampling over a single
($\Gamma$) {\bf k} point in a LUC consisting of
$2\times2\times2$ or $3\times3\times3$ single perovskite
cells (40 or 135 atoms in total, correspondingly)
and fitted to corresponding partial DOS from a LMTO
calculation. $\beta_{\mu\nu}$ affects the resonance interaction
of the ${\mu}$th AO with other states and hence
the width of the corresponding hybridized bands, which
can be as well fitted to the {\em a~priori} known partial DOS.
We used for the two-center parameter $\beta_{\mu\nu}$
a weighted value $(\beta_{\mu}+\beta_{\nu})/2$,
therefore $\beta_{\mu}$ and $\beta_{\nu}$
may be calibrated in such a case as one-center parameters.
For an initial value of $P^{(0)}$, an expected
occupation of individual AO's, based on electronegativity
considerations, in the compound in question may be taken,
and then refined in the course of iterations.
The two-center parameter $\alpha_{AB}$,
which does not depend on orbital indices,
plays a relatively minor role in what regards
the band structure and DOS and affects primarily
the energetics of atomic displacements, equilibrium
bond lengths, and hence the equilibrium geometry. Finally,
$\zeta_{\mu}$, nominally being a Slater exponent parameter
and as such tabulated for all elements, should of course
be considered here as a free parameter, which is used
to improve the quality of our fixed, single-exponent
basis set. It needs some adjustment based on a compromise
between different properties that are sought to be optimized.

\begin{figure}[hbt]
\centerline{\psfig{file=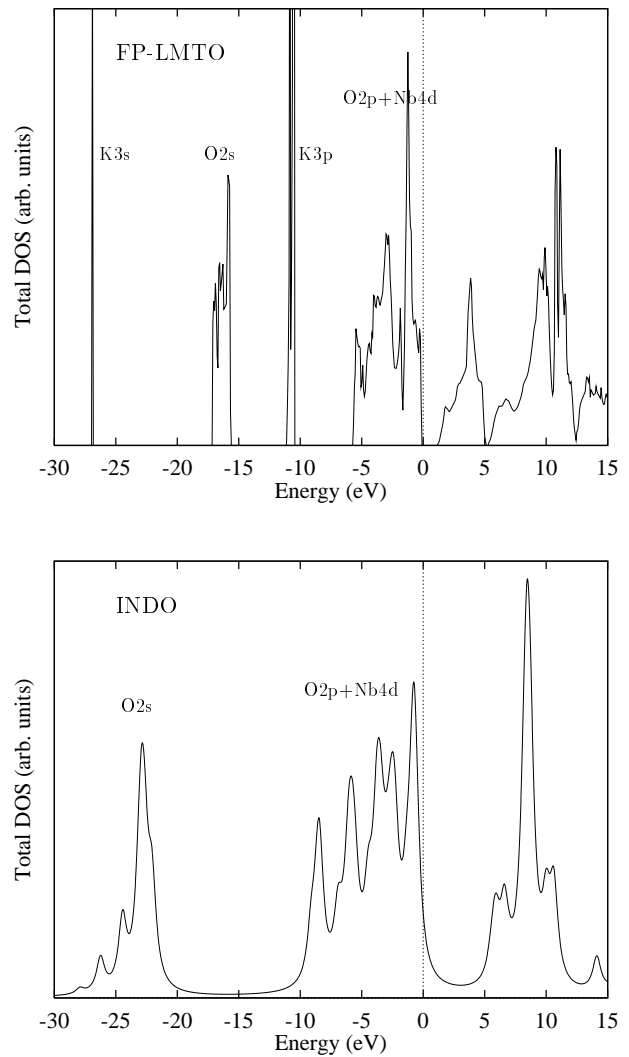,width=8.0cm}}
\caption{Total density of states of KNbO$_3$ calculated
with full-potential LMTO method (above)
and INDO method for a $3\times3\times3$ supercell (below).
Energy scale is relative to the valence band top.}
\label{fig:dos}
\end{figure}

An example of the total DOS per 135-atom LUC of KNbO$_3$
is shown in Fig.~\ref{fig:dos} along with the result
of FP-LMTO calculation. The most obvious
discrepancy is in the energy separation between the primarily
O~$2s$ band and the primarily O~$2p$+Nb~$4d$ valence band.
This difference is due to the neglect of self-interaction in
the LDA-based LMTO calculation and the lack of
correlation effects in the INDO; the experimental
x-ray photoelectron measurements set O~$2s$--O~$2p$
separation at about 15 eV,\cite{xps} halfway
between the results of Hartree--Fock and LDA calculations.
The experimental estimate of the optical gap of 3.3~eV
(Ref.~\onlinecite{gap}) is again in between the LDA value of 1.4~eV
and 6.1~eV from the INDO calculation.\cite{mark_1}
These differences have a physical foundation and cannot
be removed without attributing unreasonable values
to, e.g., O~$2s$ and O~$2p$-related INDO parameters.

\begin{table}[htb]
\caption{One-center INDO parameters}
\begin{tabular}{lcdddd}
 Orbital & \mbox{$\zeta$ (a.u.$^{-1}$)} & $E_{\text{neg}}$ (eV) &
           $-\beta$ (eV) & \mbox{$P_0$ (a.u.)} \\
\hline
   O~$2s$ &  2.27  &     4.5  &  16.0  &  1.974 \\
   O~$2p$ &  1.86  & $-$12.6  &  16.0  &  1.96  \\
  Nb~$5s$ &  2.05  &     0.0  &  30.0  &  0.1   \\
  Nb~$5p$ &  2.05  &  $-$2.0  &  30.0  &  0.0   \\
  Nb~$4d$ &  1.60  &    23.85 &  16.0  &  0.6   \\
   K~$4s$ &  1.10  &     2.8  &   2.0  &  0.1   \\
   K~$4p$ &  1.25  &     0.3  &   3.0  &  0.03  \\
\end{tabular}
\label{tab:param}
\end{table}

The effective charges found by the Mullikan population analysis
are $+0.543$ for K, $+2.019$ for Nb, and $-0.854$ for O.
This is generally in agreement with simple tight-binding
calculations,\cite{xas} but emphasizes higher degree of
covalency of the K-O bond than may be expected from intuitive
electronegativity considerations. One should note, however, that
static effective charges are not well-defined properties
and vary considerably depending on a method used.

K~$3s$ and K~$3p$ states, which were included
into the valence-band basis set in the LMTO calculation,
have been treated as core states in the INDO method.
We found that in order to obtain correct equilibrium volume,
it is essential to treat K~$3p$ states as the
basis AO's within the valence band, since their overlap
with AO's of other atoms is not negligible. This observation
is in agreement with what was earlier established in
FP-LMTO calculations.\cite{ktn3}
However, the inclusion of K~$3p$ at the expense of K~$4p$
in the minimal one-exponential basis of the INDO method
does not allow one to describe off-center displacements and
phonon frequencies with sufficient accuracy. Therefore,
we prefer to keep the K~$4p$ as a valence AO and to perform
the calculations discussed below at the experimental
lattice parameters of KNbO$_3$.

Keeping in mind the necessity to obtain reliable values
of equilibrium atomic displacements and the shape
of the potential surface related to such displacements
for subsequent studies of ferroelectric materials,
we concentrated on these values as primary criteria
for the quality of the INDO parametrization we look for.
It is known that the fine balance between long-range electrostatic
dipole-dipole interaction and the short-range chemical
bonding effects is accountable for
the ferroelectric instability, therefore the parameters
$\alpha_{AB}$ and $\beta_{\mu}$ were especially
subject to refinement, once $E_{\text{neg}}$ and $P^{(0)}$
are essentially fixed based on a band-structure analysis.
The total-energy results from the INDO calculations
for different displacement patterns
have been fitted to analogous data obtained earlier
with the FP-LMTO method as described in Ref.~\onlinecite{ktn3}.
We made sure that the optimized parameter set provides
reasonable agreement with the FP-LMTO data
in describing different displacement patterns and
is not confined to any particular symmetry.

Since the shape of the total-energy hypersurface
over atomic displacements is not directly measurable
experimentally, and the results by different {\em ab~initio}
calculation schemes differ somehow in determining the
depth and the position of the off-center potential minima
(see, e.g., Refs.~\onlinecite{sb92} and \onlinecite{s95}),
we relied also on a neutron-diffraction data concerning
the displaced atomic positions in the ferroelectric
phases of KNbO$_3$,\cite{hewat}
and on the $\Gamma$ transverse-optic (TO) phonon frequencies
as additional reference points to test our parametrization.
The one-center INDO parameters we found to provide
the best compromise in reproducing all these properties
are given in Table \ref{tab:param}.
The best-fitted two-center parameters $\alpha_{AB}$ are
0.15, 0.33 and 0.39 a.u.$^{-1}$ for $A$=O and
$B$=O, Nb and K, correspondingly, and zero for $A$=Nb and K.
The results of our ground-state geometry
and phonon calculations are discussed in the next section.

\section{Results and discussion}
\label{sec:results}

\subsection{Sequence of ferroelectric phases}

As the temperature lowers, KNbO$_3$ undergoes a sequence
of phase transitions from paraelectric cubic to
ferroelectric tetragonal then orthorhombic then rhombohedral
phases. The atomic positions in all these phases
have been determined by Hewat.\cite{hewat}
As a first approximation, each of these ferroelectric
phases is characterized by the off-center displacement
of the Nb atom from its symmetric position in the cubic
perovskite cell along [100], [110], or [111] in three
subsequent ferroelectric phases, with the gradual lowering
of the total energy. On top of this major distortion,
K and O atoms somehow adjust their positions as compatible
with the reduced symmetry of each particular phase,
and a lattice strain eventually appears. The hierarchy
of total-energy lowerings \linebreak

\begin{figure}[hbt]
\centerline{\psfig{file=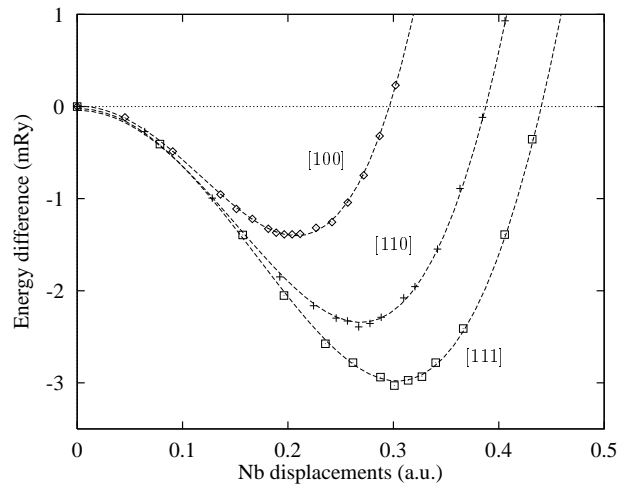,width=8.0cm}}
\caption{Total energy as a function of
off-center Nb displacements along different directions
from its position in the cubic perovskite structure
as calculated with the INDO ($2\times2\times2$ supercell) method.}
\label{fig:displ}
\end{figure}

\begin{table*}[htb]
\caption{
Calculated $\Gamma$-TO frequencies and eigenvectors in cubic KNbO$_3$.
}
\begin{tabular}{cdddddcccccccc}
 &
\multicolumn{5}{c}{Eigenvectors (present work)}& &
\multicolumn{4}{c}{$\omega$ calc. (cm$^{-1}$)} & &
\multicolumn{2}{c}{$\omega$ expt. (cm$^{-1}$)} \\
\cline{2-6} \cline{8-11} \cline{13-14}
\raisebox{2.5ex}[0pt]{Symmetry} &
  K & Nb & O & O & O & &
  Present&Ref.~\onlinecite{sb92}&Ref.~\onlinecite{phonon}&
	  Ref.~\onlinecite{loto}& &
  Ref.~\onlinecite{fmsg84}&Ref.~\onlinecite{glsq80}\\
\hline
 $T_{1u}$ &   0.05&$-$0.57& 0.70&   0.30&   0.30& &
  292$i$ &115$i$&203$i$&143$i$  & & 96    &115     \\
 $T_{1u}$ &$-$0.88&   0.34& 0.21&   0.16&   0.16& &
  178    &168   &193   &188     & &198    &207     \\
 $T_{1u}$ &$-$0.02&$-$0.19&$-$0.61&  0.54& 0.54& &
  537    &483   &459   &506     & &521    &522     \\
 $T_{2u}$ &   0.  &   0.  &  0. &  1. & $-$1.  & &
  272    &266   &234   &        & &       &280$^a$ \\
\end{tabular}
$^a$ Measurements at 585 K (in the tetragonal phase),
Ref.~\onlinecite{fmsg84}.
\label{tab:phonon}
\end{table*}

\noindent
related to the Nb displacements along three directions
is therefore an important benchmark
for the quality of the calculation in question.
In Fig.~\ref{fig:displ}, the energy gain due to the
Nb displacements from the central position in the
cubic perovskite cell (with the lattice constant
$a$=3.997~\AA) is shown as calculated by the INDO method
for the $2\times2\times2$ supercell.

As is consistent with the experimental data, the
[111] displacement and hence the rhombohedral phase
provides the lowest ground-state energy, followed
by the [110] displacement (orthorhombic phase) and
the [100] displacement (tetragonal phase).
This qualitative result is relatively stable against
some variations of the INDO parameters. As regards
the magnitudes of the off-center displacements and
the depth of the related total-energy wells, our INDO parametrization
(Table~\ref{tab:param}) provides good agreement
with the results of the FP-LMTO calculations accounting to
all three displacement directions [see Fig.~4(a)
of Ref.~\onlinecite{ktn3} and Fig.~1(b) of Ref~\onlinecite{phase}].

\subsection{$\Gamma$-TO frozen phonons in the cubic phase}

As another test for the quality of our INDO parametrization
for the adequate description of the atomic-displacement
potential surface, we calculated the $\Gamma$ TO
phonon frequencies in the cubic phase of KNbO$_3$.
Similar calculations have been done earlier by other
methods,\cite{sb92,phonon,chain} and the experimental
data (obtained mostly by infrared reflectivity
measurements\cite{fmsg84,glsq80}) are available.
We performed the calculations for a lattice constant
$a=3.997~${\AA} (that is based on an experimental perovskite
cell volume extrapolated to zero temperature) within
a conventional frozen-phonon scheme, using
the $2\times2\times2$ LUC. Consistently with
the symmetry analysis given, e.g., in Ref.~\onlinecite{phonon},
we studied the effect on the calculated total energy
of small coupled distortions compatible with the $T_{1u}$
irreducible representation, that reveals three TO frequencies,
and of the oxygen displacement within the single ``silent''
mode compatible with the $T_{2u}$ irreducible representation
of the $Pm3m$ space group. The calculated phonon frequencies
and eigenvectors are given in Table~\ref{tab:phonon}.

The calculated frequencies generally fall within the limits
set by previous {\em ab initio} calculations,\cite{sb92,phonon,loto}
with somehow better agreement for the hard $T_{1u}$ mode
and the $T_{2u}$ mode. The eigenvectors agree well with
those calculated in Ref.~\onlinecite{sb92} by the FP-LAPW
method (for the lattice constant $a=4.016$~\AA), and with
those calculated in Ref.~\onlinecite{phonon} by FP-LMTO.
Anyway, the main pattern of atomic vibrations within each
particular $T_{1u}$ mode (primarily Nb displacement in the soft mode;
almost pure K vibration against all other atoms in the intermediate
mode, and the stretching of the oxygen octahedra in the hard mode)
are correctly reproduced. The detailed structure of the soft
mode eigenvector reveals smaller participation of K in the
displacements with respect to the center of mass than was
obtained in the FP-LMTO calculation.\cite{phonon} This seems
to be consistent with the atomic coordinates in the tetragonal
phase that emerges as the soft mode freezes down
(see Ref.~\onlinecite{hewat} for the experimental data,
and Fig. 1 of Ref.~\onlinecite{phonon}), and this behavior comes out
correctly based on our INDO parametrization.

\subsection{Equilibrium displacements in the orthorhombic phase}

As an additional benchmark for the fine adjustment
of two-center INDO parameters,
we aimed at obtaining a possibly good agreement with
the experimental data\cite{hewat} in determining
all atomic positions, and not only the Nb displacement,
in the orthorhombic and rhombohedral ferroelectric phases.
The orthorhombic phase is important because it exists
in a broad temperature range around room temperature
and is subject to most studies and practical applications.
The rhombohedral phase is specially discussed below.
Keeping the lattice vectors for the orthorhombic phase
fixed and equal to those listed in Ref.~\onlinecite{hewat}
($a$=3.973 {\AA} along ${\bf x}=[100]$ of the cubic aristotype,
$b$=5.695 {\AA} along ${\bf y}=[0\bar{1}1]$ and
$c$=5.721 {\AA} along ${\bf z}=[011]$), we allowed the $c$ relaxation
of K and Nb atoms and the $b$ relaxation of those O atoms
that are in the same [001] plane with Nb in the course
of INDO calculations towards self-consistency.
The total-energy minimization is implemented in the code
making use of the downhill simplex method (see, e.g.,
Ref.~\onlinecite{numrec}).
The resulting atomic positions within the orthorhombic cell are shown
in Table~\ref{tab:equilib} in comparison with the neutron-diffraction
estimations of Ref.~\onlinecite{hewat}. It was of course our aim
to provide as good agreement as possible by an appropriate choice
of INDO parameters, but the encouraging result is that
the agreement is very good, given the small number
of parameters \linebreak

\begin{table*}[htb]
\caption{
Positions of atoms in orthorhombic and rhombohedral phases of KNbO$_3$
(in terms of lattice parameters) as determined
by neutron diffraction measurements,
Ref. \protect\onlinecite{hewat}, and optimized
in the INDO calculation.
}
\begin{tabular}{lcccrlcc}
  Atom & $a$ & $b$ & $c$ & & & ~~~~~$\Delta_{exp}$ &
			       ~~~~~$\Delta_{calc}$ \\
\hline
\multicolumn{8}{c}{Orthorhombic phase} \\
  K     & 0 & 0 & $\Delta_z$ & & &
	\begin{tabular}{d} 0.0138$\pm$71 \end{tabular} &
        \begin{tabular}{d} 0.0209        \end{tabular} \\
  Nb    & $\frac{1}{2}$ & 0 & $\frac{1}{2}$ \\
  O$_{\text{I}}$  & 0 & 0 & $\frac{1}{2}+\Delta_z$ & & &
	\begin{tabular}{d} 0.0364$\pm$10 \end{tabular} &
	\begin{tabular}{d} 0.0347        \end{tabular} \\
 \begin{tabular}{c} O$_{\text{II}}$ \\
		    O$_{\text{II}}$ \end{tabular} &
 \begin{tabular}{c} $\frac{1}{2}$   \\
		    $\frac{1}{2}$   \end{tabular} &
 \begin{tabular}{c} $\frac{1}{4}+\Delta_y$ \\
		    $\frac{3}{4}-\Delta_y$ \end{tabular} &
 \begin{tabular}{c} $\frac{1}{4}+\Delta_z$ \\
		    $\frac{1}{4}+\Delta_z$ \end{tabular} &
 $ \left. \begin{array}{c} ~ \\ ~ \end{array} \right\} $ &
 \begin{tabular}{c} $\Delta_z$~: \\
		    $\Delta_y$~:    \end{tabular} &
 \begin{tabular}{d} 0.0342$\pm$9 \\
		 $-$0.0024$\pm$9 \end{tabular} &
 \begin{tabular}{d} 0.0347       \\
		 $-$0.0028       \end{tabular} \\
\hline
\multicolumn{8}{c}{Rhombohedral phase} \\
  K     & $\Delta_z$    & $\Delta_z$    & $\Delta_z$ & & &
	\begin{tabular}{d} 0.0130$\pm$81 \end{tabular} &
	\begin{tabular}{d} 0.0139        \end{tabular}  \\
  Nb    & $\frac{1}{2}$ & $\frac{1}{2}$ & $\frac{1}{2}$ \\
  \begin{tabular}{c} O \\ O \\ O \end{tabular} &
  \begin{tabular}{c} $\frac{1}{2}+\Delta_x$ \\
		     $\frac{1}{2}+\Delta_x$ \\
		     $\Delta_z$ \end{tabular}&
  \begin{tabular}{c} $\frac{1}{2}+\Delta_x$ \\
		     $\Delta_z$ \\
		     $\frac{1}{2}+\Delta_x$ \end{tabular}&
  \begin{tabular}{c} $\Delta_z$ \\
		     $\frac{1}{2}+\Delta_x$ \\
		     $\frac{1}{2}+\Delta_x$ \end{tabular}&
  $ \left. \begin{array}{c} ~ \\ ~ \\ ~ \end{array} \right\} $ &
 \begin{tabular}{c} $\Delta_x$~: \\ $\Delta_z$~:   \end{tabular} &
 \begin{tabular}{d} 0.0301$\pm$9 \\ 0.0333$\pm$15  \end{tabular} &
 \begin{tabular}{d} 0.0213       \\ 0.0328         \end{tabular} \\
\end{tabular}
\label{tab:equilib}
\end{table*}

\noindent
that are not directly related to the structure
properties. The most noticeable discrepancy is in the
relative displacement of K atoms, which was somehow overestimated
as compared with the experimental value; the similar trend
obtained in the FP-LMTO optimization of the orthorhombic phase
was much more pronounced.\cite{ortho} One should note
that the error in determining the position by neutron scattering
is, of all atoms involved, maximal for K,\cite{hewat} and
that our present estimate falls within the error bars
given in Ref.~\onlinecite{hewat}.

\subsection{Displacements and phonons in the rhombohedral phase}

The rhombohedral phase corresponds to the low-temperature
ground-state structure of KNbO$_3$,
which is what any zero-temperature total-energy minimization
should normally drive at, with all structural constraints
lifted. We looked for optimized atomic positions compatible
with the symmetry of the rhombohedral phase, using the
lattice constants of $a$=$b$=$c$=4.016~\AA (Ref.~\onlinecite{hewat})
but keeping the rhombohedral strain angle fixed.
The reason for this was that the total energy was found
to be very insensitive to the rhombohedral strain in BaTiO$_3$
in a FP calculation by Cohen and Krakauer,\cite{cohen}
and we do not expect to achieve better accuracy
for KNbO$_3$ in our INDO calculation. Moreover, we
neglected the deviation of the rhombohedral strain angle
$\alpha=89.83^{\circ}$ from 90$^{\circ}$.

The experimental and our optimized atomic positions
in terms of lattice vectors are given in Table~\ref{tab:equilib}.
This is, to our knowledge, the first optimization
of the atomic positions in the rhombohedral phase
of KNbO$_3$. The maximal discrepancy with the experiment
is for the $\Delta_x$(O) parameter that describes
a slight stretching of oxygen octahedra. This parameter
is obviously related to the rhombohedral strain
and may be slightly adjusted in a calculation incorporating
the exact value of the strain angle.
The relative displacements of K, Nb, and O
along the polar [111] axis are all found to be
in very good agreement with the experiment.
The quality of the description of the total-energy hypersurface
in the rhombohedral phase was further controlled
by calculating the $\Gamma$ TO-phonon frequencies.
The general symmetry relations between the $\Gamma$
phonon modes in the cubic and rhombohedral phases may be found,
e.g., in Ref.~\onlinecite{fmsg84}. We consider in the
present work only three $A_1$ modes, which originate
from the $T_{1u}$ block of the cubic phase,
as the crystal symmetry lowers and all the soft modes
become stabilized. The calculated frequencies and eigenvectors
are given in Table~\ref{tab:rhom}. The components
of the eigenvector related to K and Nb displacement
exist only along [111], whereas each of three equivalent O atoms
may also have the normal component of the displacement,
in the threefold axial symmetry along the polar axis.
The experimental phonon frequency data for the rhombohedral
phase do not seem to be numerous; the values shown in Fig.~8
of Ref.~\onlinecite{fmsg84} are about 200, 270 and 600 cm$^{-1}$.
Our calculated frequencies are in good agreement with
these data. It is interesting to compare the eigenvectors
with those for the cubic structure. One can see that the
``pure K'' mode is not affected by the structure transformation,
preserving almost exactly its frequency
and the displacement pattern. The former soft mode of
the cubic phase only slightly changes the eigenvector, but
gets hardened up to 278 cm$^{-1}$ in the rhombohedral structure.
Finally, the highest-frequency mode has the lowest
contribution of K and Nb displacements and is essentially
related to the stretching of the oxygen octahedra, as in
the cubic phase.

\begin{table}[htb]
\caption{
Calculated frequencies and eigenvectors
of the $\Gamma$-$A_1$ modes in rhombohedral KNbO$_3$.
}
\begin{tabular}{cdddd}
 & \multicolumn{4}{c}{Eigenvectors} \\
\raisebox{2.5ex}[0pt]{$\omega$ (cm$^{-1}$)} &
  K$_{\parallel[111]}$ & Nb$_{\parallel[111]}$ &
  O$_{\parallel[111]}$ & O$_{\perp[111]}$ \\
\hline
 173 & 0.88 & $-$0.37 & $-$0.16 & 0.04 \\
 278 & 0.03 & $-$0.53 &    0.40 & 0.28 \\
 593 & 0.04 &    0.26 & $-$0.23 & 0.51 \\
\end{tabular}
\label{tab:rhom}
\end{table}


\section{Summary}

In applying the semiempirical INDO method
to the study of a cubic perovskite system, we demonstrated
that the method is sufficiently sensitive for the adequate
decription of a ferroelectric instability.
The energy gain of the order of $\sim$1~mRy per unit cell,
i.e., much lower than one has to deal with in other
conventional applications of the INDO method, are nevertheless
reliably reproduced, resulting in a correct description
of the microscopic structure of ferroelectric orthorhombic
and rhombohedral phases and of the $\Gamma$ TO-phonon frequencies and
eigenvectors. The choice of the INDO parameters was
proposed for the Nb-containing system,
and may be used in further applications.

\acknowledgements
The work has been done as part of the German-Israeli
joint project ``Perovskite-based solid solutions
and their properties.'' Financial support by the
Nieders\"achsische Ministerium f\"ur Wissenschaft und Kultur
and by the Deutsche Forschungsgemeinschaft
(SFB~225) is gratefully acknowledged.
The authors are grateful to Yu.~F.~Zhukovskii
and E.~A.~Kotomin for helpful discussions.


\begin{references}
\bibitem{mark_1} The gap region is smeared up in the
		 lower panel due to the broadening
		 introduced while constructing the DOS
		 by sampling.
\end{references}
\end{document}